# An Ontology-based Context Aware System for Selective Dissemination of Information in a Digital Library

Marisa R. De Giusti, Gonzalo L. Villarreal, Agustín Vosou, Juan P. Martínez

**Abstract**— Users of Institutional Repositories and Digital Libraries are known by their needs for very specific information about one or more subjects. To characterize users profiles and offer them new documents and resources is one of the main challenges of today's libraries. In this paper, a Selective Dissemination of Information service is described, which proposes an Ontology-based Context Aware system for identifying user's context (research subjects, work team, areas of interest). This system enables librarians to broaden users profiles beyond the information that users have introduced by hand (such as institution, age and language). The system requires a context retrieval layer to capture user information and behavior, and an inference engine to support context inference from many information sources (selected documents and users' queries).

————————— ◆ —————————

• Marisa Raquel De Giusti is with Comisión de Investigaciones Científicas de la Provincia de Buenos Aires (CICPBA) and Proyecto de Enlace de Bibliotecas (PrEBi).r

• Gonzalo Luján Villarreal is with Consejo Nacional de Investigaciones Científicas y Técnicas (CONICET) and Proyecto de Enlace de Bibliotecas (PrEBi).

• Agustín Vosou is with Proyecto de Enlace de Bibliotecas (PrEBi).

• Juan Pablo Martínez is with Proyecto de Enlace de Bibliotecas (PrEBi).

## 1. INTRODUCTION

UNIVERSITIES, research centers and most teaching and researching institutions hold and produce periodically loads of knowledge and information. Besides their own knowledge, these institutions are always looking for new means to access documents from the same areas kept back in other research centers. This wealth of knowledge must be properly organized and disseminated in order to maximize its use. Information and Communication Technologies (ICT) offer every day new and better tools for cataloging, storing, retrieving and distributing knowledge, and the current tendency seems to point to Institutional Repositories (IRs) or Digital Libraries (DLs), where information can be managed and users can access to new or improved services. The combination of these repositories with most currently active initiatives for open intellectual creation sharing —such as Open Access Movement (OA), Open Archives Initiative (OAI), Creative Commons and Budapest Open Access Initiative (BOAI)— results in the availability of millions of content records. One big challenge for repositories is to find which mechanisms are the most suitable for obtaining, holding and offering these records to their users.

In contrast to general purpose search engines (such as Google, Yahoo! and Bing), IRs usually offer high-quality scientific and academic documents. DLs and IRs users —researchers, professors and high level students— are characterized by their specialization in one or more areas of knowledge, and by their permanent interest for updated information in those areas. Users attention are captured in DLs and IRs by different services oriented to make it easy to publish and distribute their work, and to gain access to other research works inside the same scope.

The addition of new research works must be publicized as soon as possible so interested users can take advantage of them. Now a new challenge arises from the past definition: how to identify which users might be interested in a determined work (extended, of course, to thousands of users and millions of documents).

Users' profiles are the main source to find out which additions might be interested for them. A proper profile representation will enable the system to obtain information beyond what the user has specified. This way, it could be possible to apply inference mechanisms and comparisons against other users' profiles. For that reason, user profiles must be seen like a complex feature rather than just a list of preferences; there must be considered a whole context in which profiles exist and relate one to another. Alternatives include key-value representations, object model and ontologies. This work is focused on the use of ontologies to represent users profiles as part of a context, because of its dynamic nature to extend profiles from and adapt profiles as the context changes. Advantages and possibilities given by this representation are considered for a Selective Dissemination of Information (SDI) service in the Intellectual Creation and Dissemination Service (Servicio de Difusión de la Creación Intelectual, SeDiCI), La Plata National University (UNLP) institutional repository.

## 2. SeDiCI, UNLP institutional repository

SeDiCI was born with the purpose of socialize the knowledge generated in all academic areas of UNLP, aiming to give back to the community the efforts put to the Public University. This main purpose covers others more specific, including:

o to offer a service for digital theses, making them public to the whole local and international community and generating links among researchers;
o to create a local culture about DL use, and to encourage researchers to share their work in a common space for all disciplines;
o to include UNLP into other existing digital resources networks .

Even though SeDiCI was first created as a theses-only service, it was almost immediately extended to all sort of digital documents to satisfy users' needs from the different Schools. Given the heterogeneous nature of UNLP, SeDiCI holds a variety of documents which includes scientific papers, pictures, musical documents, conference presentations, research projects and, of course, theses.

### 2.1 Inside SeDiCI: services for users

SeDiCI users can be document authors, project directors, researchers or simply readers. All users can access all existing services inside SeDiCI. To mention a few of them, users can create folders and put there documents selected from a search. There also exist an on-line chat, from which users can obtain SeDiCI administrators help on very specific information topics. Users can also subscribe to searches and then automatically receive news about new documents added to the repository that match their query according to a free temporal scheme (every 15 days, every month, every week, etc); this can be considered as a initial scheme for a SDI service.

### 2.2 Outside SeDiCI: interfacing repositories around the world

UNLP members and external SeDiCI users always need updated contents from multiple disciplines. To achieve this purpose, SeDiCI plays the role of OAI Service Provider over an increasing amount of external repositories, exceeding 12 million information records so far. As counterpart, SeDiCI also plays the role of Data Provider, by which all works created inside UNLP and stored/published by SeDiCI are offered to any OAI Service Provider.
UNLP libraries can also access SeDiCI documents via web services, and offer a larger set of resources to their own users, in a completely transparent fashion.

## 3. SDI in dynamic web environments

In the field of DL, many systems for content-based recommendation have been designed, in which notifications are sent periodically or by request, informing users about existing resources according to their interests. This kind of service, which try to satisfy highly specialized users with very specific needs, are based on a predefined profile created in the library for each user[1] [2].
SDI is a process by which users express their information needs either explicitly or inferred by the system and then receive notifications through information providers participating in the SDI. Users profiles may take different shapes, from a text-free query, a SQL query or a set of rules.

For a successful SDI it is necessary a solid user profile configuration, consisting of a selection of languages, document types, publication years, countries of origin, authors and a notification mean. Besides, the system must make sure that every time a user receives news, he will quickly identify which recommendations correspond to which profiles (given that users may have more than one profile).

One of the biggest problem of SDI systems is the identification of users profiles. In general, users can specify certain parameters such as Preferences, Areas of Interest and probably Subscription to Searches. However there is much more contextual information —which belongs to the user profile— that users are not always able to define, either because of system limitations, complexity of the information itself, or simply because users do not always recognize all their needs. Consequently it seems evident that systems must be extended to improve or optimize users profiles, capturing information from the context, inferring data from each profile and identifying opportunities beyond users explicit preferences.

## 4. CONTEXT-AWARE DIGITAL LIBRARIES

### 4.1 The Context

There exist many definitions for context; most of them have similar characteristics for the scope of this work. The concept of context has been studied by philosophers, psychologists, linguistics and recently engineers. From the computer science point of view, context has been defined as formal, abstract and first-class free of representation citizens in Logic and Artificial Intelligence; as routines over a set of entities in Programming Languages; and as sorted set of pairs with some operations among them in Systems[3].

Reto Krummenacher y Thomas Strang [4]have defined context as "any information that can be used to characterize the situation of an entity. An entity can be a person, a place or an object considered relevant for the interaction between the user and the application, including the user and the application itself".

In the paper Context Aware Retrieval in Web-Based Collaborations[5], authors define context as "any information used to define the user environment", and they highlight the difference between Current Context (CC) and other user contexts.

In general, context definitions include both the user and the information associated directly with each user, which is the user profile. This is not too different in the field of digital libraries, since the context of the user is made up of a set of areas of interest (or research areas), the user work group members, all resources selected or downloaded by the user and even all queries made to the system.

### 4.2 Context representation through Ontologies

There exist many ways to represent the context:

- o  via an Object Model;
- o  using a markup language;
- o  with a set of key-value pairs;
- o  based on logic;
- o  using graphics.

Traditionally, the model of the context is created following a top-down mechanism: first the application and its functionality is defined, and then the necessary ontologies for the context are developed. Ontologies are commonly used to formalize taxonomies that represent types and values of simple properties. But there is more behind the ontology-based modeling. A generic and reusable ontology will have a direct impact in the interoperability of context-aware systems, and therefore will have a direct influence on the speed to create, implement and integrate new applications. A well designed model is a key factor to access the context as well as to adapt to changes, which is very common in dynamic systems.

In the work A Context Modeling Survey[6], authors identify 6 main requirements that any context model applied to ubiquitous systems must accomplish:

o Distributed Composition: ubiquitous systems derive directly from distributed systems. The composition and administration of the context model and its data has a clear dynamism that varies in terms of time, network topology and source of information.

o Partial Validation: it is desirable to validate contextual knowledge both in the structure as well as in the instance level, even if the model is not located in one single node because of its distributed nature. Given the complexity of interrelations in the context, it turns very important to validate it.

o Richness and quality of information: information sources to characterize entities are very different, but this should not affect the quality of the information.

o Lack of complexity and ambiguity: in particular, if the information is retrieved from sensors or inferred from other information sets. The model should be able to interpolate incomplete data, only if this interpolation does not affect data quality.
o Level of formality: to describe facts and relationships in a precise and traceable way.
o Applicability to existing environments: a context model must allow its use in pre-existing computing environment infrastructure.

Ontologies are a powerful tool to specify concepts and relationships. They provide formalizations to map real-life entities to computer-enabled data constructions. In this sense, ontologies provide a uniform methodology for specifying model concepts, sub-concepts, relationships, properties and facts, and all this provide the basis for context knowledge sharing and information reusing. With this information, computer applications can determine contextual compatibility, compare contextual facts, infer new facts and even new contexts. The possibility to infer new context results particularly interesting for the lack of completeness of the context.

Although the use of ontologies has many advantages, it is not always easy to differentiate them from using other methodologies. Object Oriented Models also provide class hierarchy, and therefore they permit at least a limited formalization of instances and classes dependency models. In addition to that, OO models just like ontology-based models permit to achieve the distributed composition requirement; partial validation in OO models is also possible, typically using a compiler at the structure level and an execution environment at the instance level[6]. Thus it is necessary to analyze the needs of the computing applications involved in each case. Conclusions lead to assert that an improved user experience is generally based on data providing from sensors and different information sources. This makes applications strongly bound to the context require to cope with more and more heterogeneous data (which also counts for ambiguity, quality and contextual data validation problems). OO models require, for instance, a low level implementation of these relationships to achieve interoperability and therefore they not always result proper for knowledge sharing in open and dynamic environments[7].

There exist many alternatives to represent the context through ontologies. In [8] authors find necessary to normalize and combine knowledge coming from different domains, and they propose a highly normalized and formal ontology-base model. In the year 2003 the language CoOL (Context Ontology Language, [9]) was introduced. This language, derived from ASC (Aspect-Scale-Context) model, can be used to enable context sensibility and contextual interoperability during service discovering and execution. The proposed architecture is distributed, and has among its element a core with a reasoner, able to infer conclusions about the context based on an ontology defined with CoOL. This ability to infer information from preexisting data is particularly interesting for the scope of this work, as will be seen below.

Following a similar line is CONON (Wang et al. [9]). Even though the idea is basically the same as CoOL (knowledge reuse ability, logic inference, knowledge sharing), there exists a highest level ontology which captures general characteristics from context entities, as well as a collection of sub-domain specific ontologies. CONON ontologies are represented via OWL-DL which, thanks to the use of description logic, permits consistency checking and contextual reasoning using inference engines developed for description languages.

One last and interesting approximation is represented in the CoBrA (Context Broker Architecture, [10]) system, which provides a set of ontological concepts for characterizing entities such as people, places or any object inside its context. The idea behind CoBrA is to provide additional support for agents with limited resources to make them context-aware. This is achieved through an architecture that helps these agents to acquire knowledge, to reason about that knowledge and to share it with the context. There is a Context Broker which maintains and manages a shared contextual model for a community of agents. These agents can be applications being executed from mobile devices, services provided by devices in a room and web services simulating the presence of people, places and objects in the physical world. In this work, it is set out an intelligent meeting room, focusing specially on users (such as Alice) predefined profiles, and showing how broker and agents cooperate with the system to make decisions based on the information pointed out by the user but also from inferred data from user profiles and the context of the meeting. User profiles define a set of rules and constraints, which tells the broker what user information can be share with the context, and what information from the context should arrive to the user.

### 4.3 Context Retrieval

Context-aware retrieval (CAR) is an extension of classic Information Retrieval (IR) that adds contextual information in the retrieval process, with the purpose of deliver relevant information to users in a current context [11].

Given that user's profiles take an important part in the context, information contained there results very suitable as a first filter of useful information. A basic profile allows to capture some information about the language, age, sometimes education level and probably something about family composition. More complete profiles include also information about interesting subjects for the user, careers, and current and past research and study areas.

Even though profile information results very useful, in many cases it lacks of completeness and thus its utility results limited. In the field of digital libraries, it is priority to offer to users access to documents related to their subjects of interest (journal articles, congress proceedings, books or book chapters, theses). The information that the user explicitly delivers to the system must be extended to achieve a higher quality service. It is specially important to detect users' needs even if the user has not requested them. This approximation does not depend on a particular domain, but can be adapted to any other domain since it is just a generic way to interact and access information.

In these days where information flows everywhere and users receive tons of newsletters, advertising or emails everyday, it gets highly important to show only relevant information to the users and to avoid sending useless information. This is the reason why it is so important to make the user profile as complete as possible, identifying real areas of interest. In this sense it turns absolutely necessary to develop communication strategies that highlight to the user the importance of complete as much as possible its own profile in order to help the system to be really accurate.

Context changes and therefore context updates represent another problem to CAR systems. Application context in the scope of DL can vary from at least two places:

- o users research and development subjects may get more specific, may open to a wider subject in the same area, or may even turn towards another direction different from the current working line;

- o in some areas, changes and advances in the research can make information outdated and obsolete, requiring its replacement for something newer.

A context-aware application capable of semi-automatically retrieve user information must be able to detect these kind of context changes and update objects as long as changes happen.

## 4.4 Information sources

As mentioned above, user profile is one of the main information sources, but it is definitely not the only one. Next there is a list of some possible useful information sources to make user profiles more complete:

*Research/work group and role inside that group*

All research groups have some kind of structure which may include a director, main researchers, PhD students, technical team and support team. Thus information given to each member of a group will depend not only on the research subject but also on the role inside the group. For instance, a main researcher with a team of scholarships may need access to information from himself but also from his team; but every member of that team may not be authorized to see other members' information.

*Selected and/or downloaded documents*

Digital Libraries usually let their users create a set of folders to store documents they consider somehow relevant. These documents, which belong to the library collection, are extremely useful to understand what the user is looking for or studying: a good cataloging will enable the system compare inside the library thesaurus and detect additional similar documents.

*User's queries*

Digital Libraries websites, and in particular SeDiCI website, permit many mechanisms to access digital documents. There always exists some kind of on-line search form with a set of filters, which can be a simple search expression or even a wide set of very specific options. SeDiCI use a combination of both approximations: users can enter a text-free expression, and they can also add as many filters as they need from a growing set of more than 50 so far. Users can also specify whether they want documents from a specific collection (only theses, journal papers, etc). Documents retrieved via OAI PMH are also distinguished from those that belong to the intellectual creation of UNLP (external documents).

Search features mentioned above are similar to most search engines, and their behavior can be compared to most On-line Public Access Catalog (OPAC) because of its advanced search options and complex filters.

SeDiCI users can also access all documents via the Exploration feature, where resources are classified and listed in sets, according to different criteria:

- o Subject and sub-subjects.
- o Document Type (theses, paper, dissertation...).
- o Degree, in case of theses (PhD, master, bachelor, specialist).
- o Repository (for external documents).

Information from explicit searches performed by users, as well as implicit searches (virtual tour), are very useful to infer and complete profiles: it is possible to know which areas users have surfed, what kind of documents they have looked for, or which world repositories they have been interested in.

*Crossing profile information*

This technique is currently used in many fields, such us on-line shopping. The idea is to compare profiles and try yo detect similar behavior patterns (visitors that bought that product were also interested in these products). If a user profile indicates that the user has selected or downloaded many articles from some journal written by some author, the system could offer to another user interested in a similar area these documents as additional resources.

This idea presents new challenges, in particular about how the context is designed and represented:

- *When is an area or subject compatible with another?* Knowledge areas usually overlap, and its organization is not always a top-down hierarchy. A graph structure is probably a better solution, connected with relationships such as is-sub-area-of or is-related-to, and probably a list of rules or steps to identify when an area is said to be compatible with another. Again, the use of ontologies to represent this information seems to be a viable alternative.
- *When is a document from the same author relevant?* Many researchers have lots of works and papers, but not necessary in the same area or with the same importance. Besides, some works might be similar or present repeated information.
- *How to sort documents in an appropriate hierarchy and show the most relevant ones?* A single information cross may throw hundreds or thousands of possible options. There should exist some method to sort and filter documents in order to deliver only relevant results: if the user receives more information than he can handle, he will definitely loose interest.

## 5. CONCLUSIONS

Inside SeDiCI users may find basically two main documents sets: those that belong to UNLP intellectual creation, and those retrieved from institutional repositories. While the amount of local resources is about few thousands (currently about 10 thousand), external resources are incremented much faster every year: about 300 thousand in 2004, almost 700 thousand in 2006, 3.5 million in 2008 and more than 10 million currently. This growth shows by itself how dynamic the repository can be, and how important is to keep students, professors and researchers up-to-date about new additions.

Moreover, SeDiCI offers services to other UNLP libraries as mentioned above, allowing them to integrate SeDiCI resources to their own in search results[12]. This means that there exist at least to classes of users: local users and UNLP libraries —and their own users.

On one hand SeDiCI users have at least one profile, and probably use one or more SeDiCI services. The activity of these users in the website is relatively reduced, since in general they search something, find the information they need, download the files and then carry on with their lives. SeDiCI could capture this behavior add information to the context.

On the other hand, UNLP libraries do not have a predefined profile. Instead, they request resources to SeDiCI through web services, according to certain criteria. The lack of profile can be overcome by the permanent and very active interaction of these websites and SeDiCI. This way, it is possible to characterize libraries profiles with contextual information, from queries launched from their own users against SeDiCI.

The context of every user, either individuals or libraries, will definitely belong to a broader context: the application context. On this global context it is possible to analyze from a higher level information about contexts, and make conclusions that permit to orient the direction of the repository from multiple locations: focusing in determined thematic repositories when retrieving resources, publishing news and specific information about some subjects, incorporating new functionality and services from users' needs, stressing certain sections in the website, and other possible changes. A new set of applications can be implemented to obtain additional information from users: traffic analyzers, logs data mining software and surveys seems to be the first candidate applications. With these additions new services will arise, demanding a new model to achieve them. An interesting model that has been studied some time ago in Spain and other countries used the positioning analysis model (related to culture, scientific and educational) in five well-defined planes: 1) the study of a product or service, 2) the client or user, 3) the institution sociability, 4) its web visibility and 5) the monitoring of the competition. Conclusions extracted from a new model like this one will be useful to define general lines for web production, improvement of contents and services, and the addition of new services with a secure demand.

**M. R. De Giusti** is an Engineer in Telecommunications (1980), a Literature Professor (2008, cum laude), professor of Computer Science School of University of La Plata and researcher without director at Comisión de Investigaciones Científicas de la Provincia de Buenos Aires. She is the Director of both PrEBi and SeDiCI projects since their conception. In the year 2009, she has also become the Director of the Library Linkage initiative, from the Iberoamerican Science and Technology Consortium (ISTEC). In the year 2005 SeDiCI was awarded with OEA's INELAM award, which was presented to De Giusti in Mexico DC. She has participated in several research projects in many fields, and her current research interests are focused on Digital Libraries, Information Management and Systems Simulation.

**G. L. Villarreal** holds a Computer Analyst degree (2004, cum laude) and a System Bachelor (2008, cum laude). He is also teaching Simulation and Models with De Giusti in Computer Science School. He has worked as a developer and researcher in PrEBi, and he is currently doing a Ph.D. in Computer Science in the field of Simulation and Simulation teaching.

**A. Vosou** is a student from Computer Science School of UNLP. He started working at SeDiCI in 2009 as developer and researcher, and his work is related to Semantic web and the use of ontologies in Digital Libraries.

**J. P. Martínez** is a student from Computer Science School of UNLP. He started working at SeDiCI in 2009 as developer and researcher, and his work is related Information Delivery Systems such as Selective Dissemination of Information.